# Skyrmions in CeFeB Amorphous Nanodisks

Mouad Fattouhi[1,*], Moulay Youssef El Hafidi[1], Mohamed El Hafidi[1]

[1]Condensed Matter Physics Laboratory, Hassan II University of Casablanca, Faculty of Science Ben M'sik, B. P 7955, Av. D. El Harty, 20663 Casablanca, Morocco.

**Email address:**

fattouhi.mouad-etu@etu.univh2c.ma

**Abstract:** Rare earth-Fe-B permanent magnets are usually used because of their excellent magnetic performance at room temperature. The high price of these elements pushes some researchers to find new alternatives. From rare-earth elements cerium (Ce) is considered as the most abundant and cheap, therefore, the interest was focused to study the properties of alloys based on this element in recent years. In this work, we investigate the magnetization behavior of CeFeB nanodisks in presence of Dzyaloshinskii-Moriya interaction. We show that topological structures, Neel skyrmions, in particular, could be nucleated spontaneously in this kind of system. We also study skyrmions behavior in the presence of finite temperature. We demonstrate that CeFeB nanodisks could host magnetic skyrmions at room temperature.

## 1. Introduction

Currently, magnetic skyrmions are considered from the most valuable magnetic textures. By virtue of their huge potential in future spintronic applications, they held a particular interest. Several experimental observations and theoretical studies show that skyrmions could exist in both bulk and thin ferromagnetic systems [1]. T. Skyrim achieved the first formulation of magnetic skyrmions in 1960 and they were observed experimentally for the first time at 2009 by a German research team [2, 3].

The palpable benefit of skyrmions is their good contribution to the new generation of magnetic data storage and logic devices. Their dynamical properties in moderate electrical current give them the ability to transport information easily [4]. In fact, many obstacles such as thermal noise and SkHE (Skyrmions Hall Effect) were encountered in the way to make skyrmion-based devices practically commercialized. Prodigious efforts were deployed to develop alternatives in order to solve these awkward problems [5, 6]. Diverse realization of spintronic devices, particularly in data storage was reached thanks to magnetic Skyrmion. One of the appealing non-volatile memories are the Skyrmion MRAMs (Skyrmion Magnetic Random Access Memory) [7], which are mainly based on magnetic skyrmions manipulation in spin-valve structures [8]. The magnetic skyrmion racetrack memories are also one of the relevant success realized recently [9], this kind of memories are based on the current-controlled movement of magnetic skyrmions in magnetic nanotracks [10].

In this paper, we aim to investigate the ground-state magnetization and skyrmions stability in CeFeB amorphous nanodisks. The investigated alloy 'CeFeB' is considered from the Rare earth-Fe-B permanent magnets which are characterized by their magnetic performance at room temperature. In recent work, some researchers examined the magnetic ground state in CeFeB nanodisks without DMI [11]. Our approach consists of elucidating the impact of DMI on the magnetization behavior of this alloy using micromagnetic simulations.

The manuscript will be organized as follows: In section 2, we present the micromagnetic approach and we describe the structure and geometry of the investigated alloy, then we explain our simulation conditions and we quote characteristics of used material. In section 3, we summarize and discuss the main results and in the last section, we give our conclusions and perspectives.

## 2. Simulations method

Nowadays, several computational methods are performed to investigate the magnetization behavior of magnetic nanomaterials. Among these methods, we opt-in our work to micromagnetic simulations to study the ground-state magnetization of CeFeB nanodisks. We used Mumax3 which is a GPU-accelerated micromagnetic simulation software developed by Arne Vansteenkiste & al. [12]. The software is based on the Landau–Lifshitz–Gilbert (LLG) dynamical equation and uses the finite difference method to provide results. Temperature simulations are included by means of a random field term in the LLG equation, which were suggested among the Brown-Langevin fluctuating field [13]. The magnetization equilibrium in hysteresis calculations is reached using the conjugate gradient method. The theoretical formalism used to achieve this study is developed in the appendix.

Micromagnetic calculations were performed in specific conditions, the simulation procedure will be provided later. Our film is cylindrical with parameters "d" and "t" quoting respectively the diameter (along the x-axis or y-axis) and the thickness along the (z-axis). The disk diameter "d" was fixed in 200 nm whereas disk thickness "t" was chosen to be 1.5nm. The initial magnetization was a thermally neutralized state (random).

Magnetic parameters of $Ce_{20}Fe_{14}B_{66}$ amorphous alloy at room temperature was taken as follows [11]: saturation magnetization $M_s=9.31\times10^5$ A/m, exchange stiffness $A_{ex}=5\times10^{-12}$ J/m and the damping constant α=0.05. The cell sizes were chosen $2.5\times2.5\times1.5$ nm$^3$ in order to be less than the exchange length of CeFeB $l_{ex} = \sqrt{\frac{2 A_{ex}}{\mu_0 M_s^2}} = 3\ nm$.

The interfacial Dzyaloshinskii-Moriya interaction and uniaxial anisotropy constants have been varied during simulations, to show their impact on the magnetization ground state. Simulations are performed also to study the effect of temperature on magnetic skyrmions stability in the investigated samples.

## 3. Results and discussions

### 3.1. Ground-state magnetization

Our study of magnetization ground state revealed various situations as a function of magnetocrystalline anisotropy and interfacial DMI as shown in Fig.1. The magnetization profiles presented in Fig.1 was obtained by performing calculations for each value of iDMI and $K_{mc}$. Let us discuss the magnetizations found by our simulations. To start, we have three noticeable situations from the figure: Multidomain, vortex, and out-of-plane configurations.

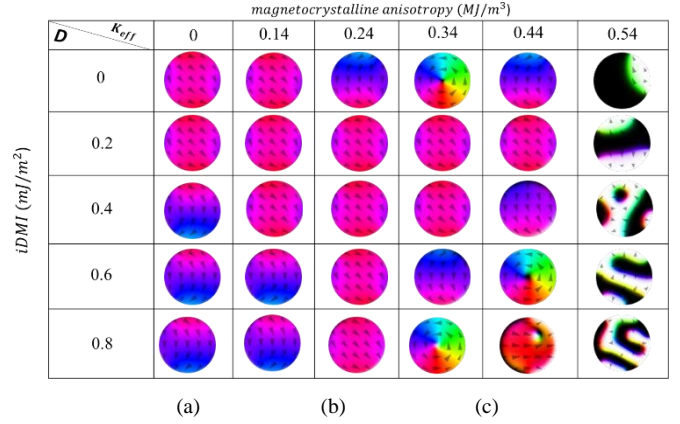

*Fig.1: Magnetization configurations of CeFeB (1.5 nm) nanodisk in the plane of iDMI and Keff.*

Magnetization profiles found in various situations are different due to various interactions interplaying on the system. In situation where we have weak values of magnetocrystalline anisotropy, a good competition between exchange interaction and demagnetization or an existence of Dzyaloshinskii-Moriya interaction can lead to vortex emergence [20]. In the situation where exchange interaction is important, we have a monodomain states and in situations where magnetocrystalline anisotropy reaches important values, we have an out-of-plan configurations where skyrmions could spontaneously have a nucleation according to the values of iDMI as shown in the figure. We notice that when anisotropy constant is between 0.3 and 0.45 MJ/m3, vortex state is favorable for some values of iDMI; however, when magnetocrystalline anisotropy constant is close to dipolar interaction, skyrmion state becomes favorable for middle values of iDMI. Let us now discuss the nature of skyrmions found in several situations. On one hand, Neel-type skyrmions could appear either in the case of ultrathin nanodisks or in ultrathin films marked by a significant Dzyaloshinskii- Moriya interaction taking place at the interfaces [14]. On the other hand, Bloch walls emerge almost in thicker films and bulk materials [15]. Skyrmions are displayed for moderate iDMI strengths in this alloy that reflect the possibility to explore such materials in experiments. Skyrmions found for high iDMI strengths have a small radius as shown clearly in Fig.2, where we reported on the out-of-plane reduced magnetization component as a function of skyrmion radius for several iDMI values. Such conditions are allowed in magnetic structures whose the space inversion symmetry is broken as in confined geometries.

In order to verify all magnetization ground-state possibilities, we performed calculations for different initial magnetization configurations (Neel skyrmion, Bloch skyrmion and Uniform).

The known behavior of domain-walls in presence of Dzyaloshinskii-Moriya interaction is well observed in our results. Bloch skyrmions (BS) are stabilized when iDMI strength is very weak however Neel skyrmions (NS) are stabilized for important iDMI strength values. Chiral multidomain (CM) state appear for high iDMI strengths (>1).

When the initial configuration is uniform, merons (Me) appeared in some situations. Table.1 contains all equilibrium states for each initial situation.

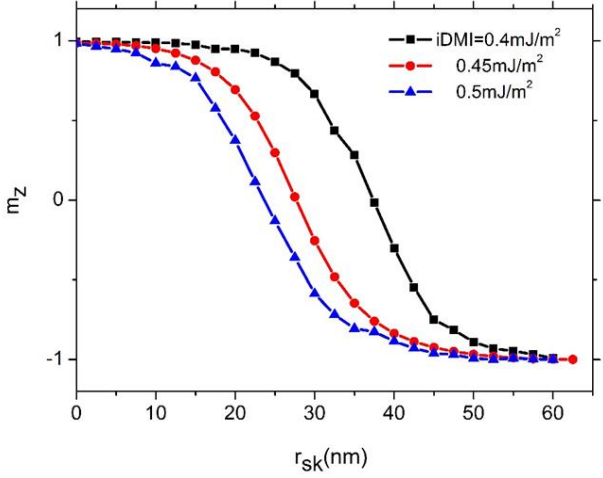

*Fig.2: Out of plane reduced magnetization component of skyrmions found in CeFeB nanodisks for Kmc= 0.55 MJ/m$^3$.*

*a)*

| Initial configuration | iDMI strength(mJ/m$^2$) | Magnetization profile |
|---|---|---|
| | 0 | U-oop |
| | 0.1 | BS |
| | 0.2 | NS |
| | 0.3 | NS |
| Neel skyrmion | 0.4 | NS |
| | 0.5 | NS |
| | 0.6 | NS |
| | 0.8 | NS |
| | 1 | CM |
| | 1.5 | CM |
| | 2 | CM |

*b)*

| Initial configuration | iDMI strength(mJ/m$^2$) | Magnetization profile |
|---|---|---|
| | 0 | BS |
| | 0.1 | BS |
| | 0.2 | BS |
| | 0.4 | NS |
| Bloch skyrmion | 0.6 | NS |
| | 0.8 | NS |
| | 1 | NS |
| | 1.5 | CM |
| | 2 | CM |

*c)*

| Initial configuration | iDMI strength(mJ/m$^2$) | Magnetization profile |
|---|---|---|
| | 0 | U-oop |
| | 0.1 | Me |
| | 0.2 | Me |
| Uniform out-of-plane (U-oop) | 0.4 | NS |
| | 0.6 | NS |
| | 0.8 | NS |
| | 1 | NS |
| | 1.5 | CM |
| | 2 | CM |

*Table.1: Magnetization ground state for different initial configuration (Kmc =0.55 MJ/m3) a); (Kmc =0.55 MJ/m3) b); (Kmc =0.45 MJ/m3) c). Bloch skyrmion (BS); Neel skyrmion (NS); chiral multidomain (CM); Meron (Me); Uniform out-of-plane (U-oop).*

### 3.2. Skyrmions stability

One of the most challenging facts actually is to stabilize magnetic skyrmions at suitable conditions to make them accessible for realistic applications. In recent works, some researchers have managed to stabilize skyrmions at room temperature by efficient experiments in CoFeB [16], Co [17]. In this section, we performed micromagnetic simulations to study skyrmions stability at 4.2 K, 100K and 300K temperatures in order to test skyrmions robustness at room temperature. The RT-FeB permanent magnets are known for their performance at room temperature. Thus, the curie temperature of the CeFeB is around 424 K in literature [18]. The results found are shown in Fig.3 where we expose the z-component of reduced magnetization as a function of skyrmion radius for different temperatures. The curves were plotted for iDMI values experimentally realizable [19].

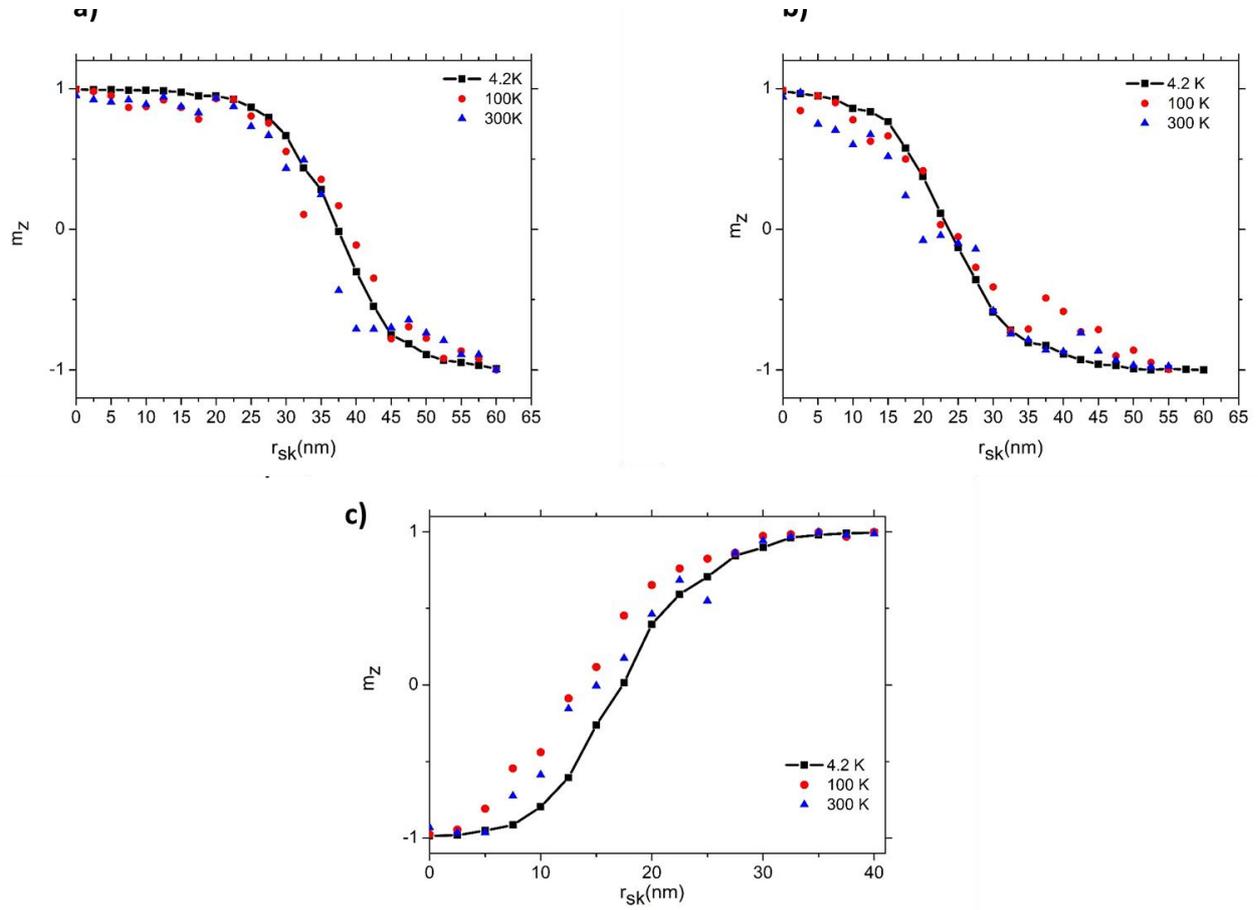

*Fig.3: Effect of temperature on skyrmions reduced magnetization out of plane component. a) situation where iDMI equal to 0.4 mJ/m², b) 0.5 mJ/m² and c) 0.7 mJ/m².*

Fig.3 shows the effect of temperature on skyrmions found in different situations: fig.3 a) is related to the situation where iDMI was 0.4 mJ/m², fig.2 b) to 0.5 mJ/m² and fig.2 c) 0.7 mJ/m². It is noticeable from the curves that skyrmions are more robust against temperature when iDMI strength is important, which is related to the competition between different energies. We remark that some fluctuations arise in the curve for high temperatures that is expected to thermal fluctuations. We can see that skyrmions are metastable for middle iDMI values that reflect the performance of the material chosen. Note that the different polarity of skyrmions found for different iDMI is due to the initial magnetization used in simulations [21].

In this section, we investigated magnetization ground state of CeFeB nanodisks in presence of Dzyaloshinskii-Moriya interaction in order to elucidate its effect. We also studied magnetic skyrmions nucleation and stability in these nanodisks.

## 4. Conclusion

In this manuscript, we studied numerically the nucleation and stability of Neel magnetic skyrmions under finite temperature and magnetic field in ultrathin CeFeB nanodisks. We investigated the effect of interfacial Dzyaloshinskii-Moriya interaction on magnetization behavior of the system. The obtained results show that skyrmions have been spontaneously nucleated in the CeFeB nanodisks for specific iDMI and $K_{eff}$ values. This work gives a clear idea about the suitable conditions for topological structures nucleation in CeFeB amorphous nanodisks.

## Appendix

In the micromagnetic theory, our system energy is given by:

$$E_M = \int A_{exch}(\nabla \cdot \boldsymbol{m}(r))^2 \, d^3r + \int D \, \boldsymbol{m} \cdot (\nabla \times \boldsymbol{m}) \, d^3r + \int d^3r \, K_{eff}(1 - \cos^2(\theta)) +$$
$$\left(-\frac{\mu_0 M_S}{2}\right) \int \boldsymbol{m}(r) \cdot \boldsymbol{H_d} \, d^3r +$$
$$(-\mu_0 M_S) \int \boldsymbol{m}(r) \cdot \boldsymbol{H_{ext}} \, d^3r \quad (1)$$

where the first term is the symmetric exchange, the second is the Dzyaloshinskii-Moriya contribution, the third is the magneto-crystalline anisotropy, the fourth is the shape anisotropy part and the last is the external field energy.

The effective anisotropy constant writes:

$$K_{eff} = K_1 + \frac{2K_S}{t} \quad (2)$$

where $K_1$ is the bulk magneto-crystalline anisotropy constant, $K_s$ is the surface anisotropy and t is the ferromagnetic layer thickness, $M_S$ is the saturation magnetization and $\mu_0$ is the vacuum permeability.

The used geometry is marked by the strength of demagnetization field "shape anisotropy" tending to keep magnetization in the plane of the sample, which is expressed by:

$$\boldsymbol{H_d} = -[N] \, \boldsymbol{M} \quad (3)$$

where $[N]$ is the demagnetizing tensor reduced for the present nanodisk to:

$$[N] = \begin{pmatrix} 0 & 0 & 0 \\ 0 & 0 & 0 \\ 0 & 0 & 1 \end{pmatrix} \quad (4)$$

The magnetization configurations are obtained by the Landau-Lifshitz-Gilbert (LLG) dynamic equation:

$$\frac{d\boldsymbol{M}}{dt} = -\gamma \boldsymbol{M} \times \boldsymbol{H_{eff}} + \frac{\alpha}{M_S} \boldsymbol{M} \times \frac{d\boldsymbol{M}}{dt} \quad (5)$$

where $\gamma$ is the gyromagnetic ratio, $\alpha$ is the dimensionless damping parameter and $\boldsymbol{H}_{eff}$ is the effective field, which is given by:

$$\boldsymbol{H_{eff}} = \frac{2A_{exch}}{\mu_0 M_S^2} \nabla^2 \boldsymbol{M} - \boldsymbol{H_d} + \frac{2D}{\mu_0 M_S^2} \boldsymbol{M} \cdot (\nabla \times \boldsymbol{M}) + \frac{2K_{eff}}{\mu_0 M_S^2} (\boldsymbol{M} \cdot \boldsymbol{e_z})\boldsymbol{e_z} - \boldsymbol{H_{ext}} \quad (6)$$

The effective field is the functional derivative of the energy density.